# A SWAP Gate for Spin Qubits in Silicon

Ming Ni,[1,2,#] Rong-Long Ma,[1,2,#] Zhen-Zhen Kong,[3] Xiao Xue,[4] Sheng-Kai Zhu,[1,2] Chu Wang,[1,2] Ao-Ran Li,[1,2] Ning Chu,[1,2] Wei-Zhu Liao,[1,2] Gang Cao,[1,2,8] Gui-Lei Wang,[3,7,8,*] Guang-Can Guo,[1,2] Xuedong Hu,[5] Hong-Wen Jiang,[6] Hai-Ou Li,[1,2,8,*] and Guo-Ping Guo[1,2,8,9,*]

[1] *CAS Key Laboratory of Quantum Information, University of Science and Technology of China, Hefei, Anhui 230026, China*

[2] *CAS Center for Excellence in Quantum Information and Quantum Physics, University of Science and Technology of China, Hefei, Anhui 230026, China*

[3] *Integrated Circuit Advanced Process R&D Center, Institute of Microelectronics, Chinese Academy of Sciences, Beijing 100029, P. R. China*

[4] *QuTech and Kavli Institute of Nanoscience, Delft University of Technology, Delft 2628 CJ, The Netherlands*

[5] *Department of Physics, University at Buffalo, SUNY, Buffalo, New York 14260, USA*

[6] *Department of Physics and Astronomy, University of California, Los Angeles, California 90095, USA*

[7] *Beijing Superstring Academy of Memory Technology, Beijing 100176, China*

[8] *Hefei National Laboratory, University of Science and Technology of China, Hefei 230088, China*

[9] *Origin Quantum Computing Company Limited, Hefei, Anhui 230026, China*

[#] These authors contributed equally to this work.
[*] Corresponding authors: guilei.wang@bjsamt.org.cn; haiouli@ustc.edu.cn; gpguo@ustc.edu.cn.

**With one- and two-qubit gate fidelities approaching the fault-tolerance threshold for spin qubits in Si, how to scale up the architecture and make large arrays of spin qubits become the more pressing challenges. In a scaled-up structure, qubit-to-qubit connectivity has crucial impact on gate counts of quantum error correction and general quantum algorithms. In our toolbox of quantum gates for spin qubits, SWAP gate is quite versatile: it can help solve the connectivity problem by realizing both short- and long-range spin state transfer, and act as a basic two-qubit gate, which can reduce quantum circuit depth when combined with other two-qubit gates. However, for spin qubits in silicon quantum dots, high fidelity SWAP gates have not been demonstrated due to the requirements of large circuit bandwidth and a highly adjustable ratio between the strength of the exchange coupling $J$ and the Zeeman energy difference $\Delta E_z$. Here we demonstrate a fast SWAP gate with a duration of ~25 ns based on quantum dots in isotopically enriched silicon, with a highly adjustable ratio between $J$ and $\Delta E_z$, for over two**

orders of magnitude in our device. We are also able to calibrate the single-qubit local phases during the SWAP gate by incorporating single-qubit gates in our circuit. By independently reading out the qubits, we probe the anti-correlations between the two spins, estimate the operation fidelity and analyze the dominant error sources for our SWAP gate. These results pave the way for high fidelity SWAP gates, and processes based on them, such as quantum communication on chip and quantum simulation by engineering the Heisenberg Hamiltonian in silicon.

**Main text.**

Single-spin qubit in a silicon quantum dot is an attractive platform for quantum information processing due to its long coherence times [1-4], robustness against thermal noise [5,6], and potential for scalability with advanced semiconductor manufacturing technology [7-9]. Rapid progress over the past years have seen high fidelity initialization [10-13], readout [12-15], single-qubit operation (fidelity above 99.9%) [12,16,17] two-qubit operation (fidelity above 99.5%) [18-21], and three-qubit iToffoli gate [22], successfully achieved, putting single-spin qubits at the threshold of demonstrating quantum error correction [22] and the noisy intermediate-scale quantum (NISQ) era [23].

For a fault-tolerant and scalable quantum computer architecture, efficient quantum information transfer/communication on chip is essential. For a single-spin qubit in Si, the transfer of qubit state can be realized by shuttling the electron directly [24-27]. The SWAP gate, via the isotropic Heisenberg exchange coupling native between quantum dots, can coherently swap states of two neighboring qubits [28], and is a natural conduit for quantum information transfer as suggested in the original spin qubit proposal [29]. Meanwhile, the $\sqrt{\text{SWAP}}$ gate allows two-qubit entangling operations and an extra SWAP gate can reduce circuit depth compared to using only the controlled-phase (CPhase) gate [30]. In short, for a large-scale spin qubit array, SWAP gate is a crucial element for both quantum computing in general and on-chip communication in particular.

Although exchange control of spins has been demonstrated in GaAs quantum dots [31] and coupled donors [32], demonstration of the SWAP gate remains a significant technical challenge for single-spin qubits in Si quantum dots due to the common presence of micromagnets that enables single-qubit gates but limit the tunability of the ratio between the exchange coupling $J$ and the Zeeman energy difference $\Delta E_z$ [28,33-36]. While an equivalent to SWAP gate can be constructed in a non-uniform magnetic field, such as the resonant SWAP gate, this more complex gate suffers from cross-talk [37] and heating, which significantly impact its applicability in large-scale qubit arrays. A recent work has demonstrated SWAP operation on a silicon double quantum dot [36], though with only limited spin flip probability due to the finite tunability of $J/\Delta E_z$.

In this article, we experimentally demonstrate a SWAP gate in an isotopically enriched silicon double quantum dot (DQD) with a built-in micromagnet, and make suggestions on how to further enhance gate fidelity. Our experiment is based on the fact that in certain regime of the detuning $\varepsilon$ between two quantum dots, the exchange splitting $J$ increases/decreases rapidly while $\Delta E_z$ decreases/increases because of the spin-orbit coupling (SOC) [38]. These opposite trends enable us to quickly adjust the ratio $J/\Delta E_z$ over more than two orders of magnitude and demonstrate a SWAP gate of high fidelity in theory (practically the gate fidelity is limited by the spin preparation and measurement (SPAM) errors in our case). Using single-qubit gates, we are also able to calibrate local single-qubit phase shifts generated during the SWAP gate in our experiment, and provide a strategy to eliminate them completely. Utilizing independent spin readouts, we estimate the lower bound of the SWAP gate fidelity and analyze the main error sources with a quasi-static noise model. With these analysis, we estimate the actual gate fidelity in our experiment, which indicates the feasibility of demonstrating a SWAP gate with a fidelity surpassing the fault-tolerance threshold.

## Results

**Measurement techniques and energy spectroscopy.** Our DQD two-qubit device is fabricated on a 70 nm $^{28}$Si epilayer with a residual $^{29}$Si concentration of 60 ppm. The device structure, as shown in Fig. 1a and 1b, is identical to the one we used in ref. [39-41] (see Methods). The complete measurement setup is described in Supplementary Information Note 1. The detuning $\varepsilon$ between the left dot $Q_L$ and the right dot $Q_R$ can be modified by adjusting the gate voltages of the plunger gates PL and PR. Here we define the detuning $\varepsilon$ at the symmetric operation point as zero, which is located at point c in Fig. 1c. The tunnel coupling $t_c$ is modified by the barrier gate BM. During the experiments, the single-electron transistor (SET) monitors the charge occupation of the quantum dots. An external magnetic field $B_0$=605 mT is applied perpendicular to the inter-dot axis, as shown in Fig. 1b, and a rectangular micro-magnet on the top of the device provides a magnetic field gradient to enable electric dipole spin resonance (EDSR) [42].

Figure 1c shows the charge stability diagram of the DQD with charge occupancy ($N_L$, $N_R$), in which charge occupancy is detected by measuring the SET conductance $dI_s/dV$. The wiggles of the boundary of (3,1) region is due to the slow tunnel rate between the electrons in the right quantum dot and the reservoir [43]. The pulsing path of the readout and initialization process is illustrated by the white arrows and colored dots in Fig. 1c and 1d. In our device, the tunnel rate of $Q_L$ to the reservoir is much faster than that of $Q_R$. Thus, for $Q_L$, the initialization and readout can be realized by spin-selective tunneling [44], which leaves the qubit in the $|\downarrow\rangle$ state, while such tunneling is not allowed for $Q_R$. To read the spin state of $Q_R$, we shuttle the electron from the right dot to the left dot followed by spin-selective tunneling [45]. After the electron spin state is read out, a spin-down electron will occupy the left dot, and we shuttle it back to the right dot to initialize $Q_R$.

In this work we focus on the (3,1) - (2,2) region. The two extra electrons in the ground orbital states in the left quantum dot ensure that the tunneling rates of $Q_R$ (to both the reservoir and $Q_L$) are appropriate for both electron counting and two-qubit operations. With two inner-shell electrons in the left dot remain inert, the low-energy dynamics of the DQD is equivalent to that in the (1,1) - (0,2) regime. Such an effective two-electron system is described by the low-energy Hamiltonian [46] in the two-spin basis of $(|\uparrow\uparrow\rangle, |\uparrow\downarrow\rangle, |\downarrow\uparrow\rangle, |\downarrow\downarrow\rangle)^T$:

$$H = \begin{pmatrix} E_z & & & \\ & \frac{\Delta E_z - J}{2} & \frac{J}{2} & \\ & \frac{J}{2} & \frac{-\Delta E_z - J}{2} & \\ & & & -E_z \end{pmatrix},$$

where $E_z$ and $\Delta E_z$ are the average and difference in Zeeman energy of the two qubits. A SWAP gate is given by the following unitary operator when $J \gg \Delta E_z$:

$$U_{SWAP}(t) = \begin{pmatrix} 1 & & & \\ & \cos(\frac{\Delta E t}{2}) & \sin(\frac{\Delta E t}{2}) & \\ & \sin(\frac{\Delta E t}{2}) & \cos(\frac{\Delta E t}{2}) & \\ & & & 1 \end{pmatrix},$$

where $\Delta E = \sqrt{\Delta E_z^2 + J^2}$ is the energy difference between the antiparallel eigenstates $|\widetilde{\downarrow\uparrow}\rangle$ and $|\widetilde{\uparrow\downarrow}\rangle$ of $H$. The ratio $J/\Delta E_z$ determines whether it is actually a SWAP gate or a CPhase gate [47,48]. A diabatic

pulse in regime $J \ll \Delta E_z$ or an adiabatic pulse would have led to a CPhase gate, while a diabatic pulse in the regime $J \gg \Delta E_z$ gives a SWAP gate [48].

**DQD Spectrum.** We first measure the energy spectrum of the two-qubit system. This is achieved by pulsing the DQD to the symmetric operation point after preparing the system in the $|\downarrow\downarrow\rangle$ state, then increasing the detuning $\varepsilon$ diabaticly. At various detuning $\varepsilon$, a frequency-chirped microwave pulse ($\pm 1$ MHz around the center frequency, $500$ μs duration time) with variable frequency adiabatically flips one of the spins. After that, the qubits are diabaticly pulsed back to the symmetric operation point and finally the spin-up probability of $Q_L$ is read-out at point d in Fig. 1c. With increased detuning $\varepsilon$, we plot the spin-up probability of $Q_L$ as a function of the microwave frequency in the bottom half of Fig. 1e. Two resonance peaks corresponding to the $|\downarrow\downarrow\rangle$ to $|\widetilde{\downarrow\uparrow}\rangle$ and $|\downarrow\downarrow\rangle$ to $|\widetilde{\uparrow\downarrow}\rangle$ transitions appear when the two qubits are in the $|\downarrow\downarrow\rangle$ state at the beginning. We attribute this uncommon phenomenon to two reasons. When the detuning $\varepsilon$ is small, the dominant reason is the single-shot readout cross-talk [34], which originates from the tunnel coupling between dots at point d in Fig. 1c. With increased detuning $\varepsilon$, $|\downarrow\uparrow\rangle$ and $|\uparrow\downarrow\rangle$ states would always mix, such that transitions from the $|\downarrow\downarrow\rangle$ state to both $|\widetilde{\downarrow\uparrow}\rangle$ and $|\widetilde{\uparrow\downarrow}\rangle$ states can be detected by reading just $Q_L$. By dividing the resonance frequency of the two peaks, we obtain the difference $\Delta E(\varepsilon)$ between the energies of $|\widetilde{\downarrow\uparrow}\rangle$ and $|\widetilde{\uparrow\downarrow}\rangle$ as a function of detuning $\varepsilon$. We can then prepare the qubits to $|\widetilde{\downarrow\uparrow}\rangle$ with the extracted frequency, and measure the resonance frequency corresponding to the $|\widetilde{\downarrow\uparrow}\rangle - |\uparrow\uparrow\rangle$ transition while increasing the detuning $\varepsilon$. The result is shown in the top half of Fig. 1e. Interestingly, the $|\downarrow\downarrow\rangle - |\widetilde{\uparrow\downarrow}\rangle$ transition remains visible here, which is most likely caused by poor initialization fidelity near the anti-crossing. With the extracted resonance frequencies, we obtain the energy-level structure in Fig. 2a, and map the exchange interaction $J(\varepsilon)$, as well as $\Delta E$ and $\Delta E_z$, at different detuning $\varepsilon$, as shown in Fig. 2b. Notice that here $\Delta E(\varepsilon)$ and $J(\varepsilon)$ are directly measured, while the Zeeman energy difference is calculated from $\Delta E_z(\varepsilon) = \sqrt{\Delta E(\varepsilon)^2 - J(\varepsilon)^2}$.

The DQD spectrum can be conveniently tuned electrically, which indicates the possibility of high tunability of $J/\Delta E_z$. When the detuning $\varepsilon$ is varied in the DQD by modifying gate voltage on each plunger gate, the electric field in the vicinity of the electron wavefunction also changes, which in turn modifies the qubit frequencies through an effective Stark shift: Due to the intrinsic spin-orbit coupling (ISOC), the vertical electric field ($F_y$) can change the qubit frequency by modifying the g-factor [1,49-51], while the electric field in the x-z plane changes the position of the electron wavefunction and induces a qubit frequency shift through the synthetic spin-orbit coupling (SSOC) [39,52]. The effects of ISOC and SSOC are comparative, both of which are dependent on the quantum dot position (see Supplementary Information Note 6). In our device, an increasing $\varepsilon$ leads to a decrease in the Zeeman energy of $Q_L$ but an increase in the Zeeman energy of $Q_R$, causing $\Delta E_z$ to decrease (see inset in Fig. 2b). Meanwhile, with the increasing $\varepsilon$, $J$ increases rapidly, which can also be tuned by the tunnel coupling $t_c$ through the barrier gate voltage. From the energy spectrum in Fig. 1e, we estimate $t_c = 1.5$ GHz and the Stark shift near the symmetric operation point to be 128.6 MHz/V. The lever arm along the detuning axis is 0.14 meV/mV, which is extracted from the magneto-spectroscopy. Numerical fitting for both $\Delta E_z$ and $J$ are given in Fig. 2b. With the parameters in Fig. 2b, the ratio $J/|\Delta E_z|$ and $J/\Delta E$ as functions of detuning $\varepsilon$ are given in Fig. 2c. With increasing detuning $\varepsilon$, $J/|\Delta E_z|$ increases rapidly and reaches a sweet spot when $\Delta E_z$ equals zero, near which $J \gg \Delta E_z$ is easily satisfied. In our experiment, we find that ratio $J/|\Delta E_z|$ can be modified by $\varepsilon$ from 0.04 to 8.4, which is more than two orders of magnitude. In reality, the adjustability of the ratio $J/|\Delta E_z|$ should be even higher, as shown by the fitting. However, we did not observe a higher $J/|\Delta E_z|$ in our experiment because of the relatively large scan step of $\varepsilon$ as well as numerical inaccuracy in $\Delta E_z$ as it approaches 0.

The highly adjustable $J/\Delta E_z$ ratio makes it possible to perform single-qubit operations and two-qubit operations simultaneously. In Fig. 2a, at the symmetric operation point $\varepsilon = 0$, the exchange interaction is small and $J \ll \Delta E_z$, so we can initialize the qubits and perform single-qubit gates. When $\Delta E_z$ is near zero, as illustrated in insert i of Fig. 2a, $J \gg \Delta E_z$, which is the perfect condition (thus we call it an operation sweet spot) for the SWAP gate. In other words, in our system, turning $\varepsilon$ can adjust $J/\Delta E_z$ significantly, benefiting from the operation sweet spot, such that the realization of the SWAP gate becomes possible in the presence of a micromagnet. To demonstrate a high-fidelity SWAP gate benefiting from the operation sweet spot, an implicit assumption here is that $\Delta E_z$ has to be smaller than a few tens of MHz at the symmetric operation point due to the limited adjusting ability of the effective Stark shift. Recall that the Stark shift has contributions from both the ISOC and SSOC. To increase the tunable range of the Stark shift, for the ISOC part, the $\frac{dg}{dF_y}$ for the two qubits should have the same signs and be as large as possible. These can be realized by altering the valley state and improving the interface smoothness [49,50]. And for the SSOC part, the effect is strongly dependent on the longitude magnetic field gradient $b_{long}$ and the electron wavefunction position, which is hard to control in silicon quantum dots. Hence, a smaller $b_{long}$ is preferred, which has the added benefit of improving single-qubit coherence [16,53].

**Exchange oscillation.** With a clear understanding of the energy spectrum of our DQD, we are ready to explore two-electron dynamics of the system. First, we demonstrate the exchange oscillation, which underlies the SWAP gate, as shown in Fig. 2d. After initializing the system to the $|\downarrow\downarrow\rangle$ state at operation point a and b in Fig. 1c, an adiabatic microwave pulse with a 2 MHz chirp modulation [54] flips the state of $Q_L$ at operation point c. As illustrated in Fig. 2a inset iii, when $\Delta E_Z \gg J$, $|\uparrow\downarrow\rangle$ and $|\downarrow\uparrow\rangle$ are the eigenstates. Hence, the two qubits will be driven to the $|\uparrow\downarrow\rangle$ state at point c. After these single-qubit preparations, pre-distorted diabatic pulses[55] on gates PL and PR rapidly increase the detuning $\varepsilon$, which brings the system to the regime where $\Delta E_Z \ll J$ while still maintaining the two-qubit state in $|\uparrow\downarrow\rangle$. When $\Delta E_Z \ll J$, the eigenstates of the two-qubit system are the singlet-triplet $(S - T_0)$ states and the spin states of the two qubits will exchange with each other at a frequency $f_{ex} = \Delta E$ (Fig. 2a inset iii). In other words, an oscillation will occur between the $|\uparrow\downarrow\rangle$ and $|\downarrow\uparrow\rangle$ states. Finally, the system is pulsed diabatically back to the region where $\Delta E_z \gg J$, and the states of both qubits are read out respectively. The exchange oscillation between the two antiparallel states as a function of $\varepsilon$ and the pulse duration $\tau_S$ is shown in Fig. 2d. We simulate the process by solving the time-dependent Schrödinger equation with $J(\varepsilon)$ and $\Delta E_z(\varepsilon)$ extracted from the energy spectroscopy. The simulated exchange oscillation fringe shown in Fig. 2e fits with the experimental results quite well.

**Local Phase Calibration.** During the exchange oscillation for the two qubits, single-qubit local phases are also inevitably accumulating. The extra phases on each qubit could have dire consequences in a quantum circuit if not properly corrected. There are two types of local phases. One part is 'axis-dependent': it only depends on the starting moment of the SWAP gate and is independent of the gate duration, which makes the rotation axis of the SWAP gate rotate in the x-y plane. Another part is 'duration-dependent': it is independent of the start moment of the gate and only depends on the pulse duration, which brings a Z-phase accumulation on each qubit. In the implementations of a CPhase gate or a CROT gate, a common approach to compensate for the local phase, which depends on the gate duration, is to apply a virtual Z-gate [46,56]. However, for the SWAP gate, only the duration-dependent part of the local phases can be compensated this way. To understand this point we first shift to the rotating frame by performing the unitary transformation $R_{logical} = \exp[-i(\hat{S}_{zL}\omega_{zL} + \hat{S}_{zR}\omega_{zR})]$, where $\omega_{zL}$ and $\omega_{zR}$ are the Zeeman energies of $Q_L$ and $Q_R$ at the symmetric operation point, respectively. Here we have taken $h = 1$. Since the two-qubit XX and YY operators do not commute with the single qubit Z operator, an extra local phase develops based on the starting moment of the SWAP operation (but does not depend on the duration of the SWAP gate) [57]. This part of the local phase is determined by the rotation axis of the SWAP operation in the antiparallel state subspace and will accumulate at the frequency of $\Delta E_z$ (see Supplementary Information Note 3) at the symmetric operation point. To eliminate this axis-dependent local phase, the idle time between the start time of the first single qubit gate and the start time of the SWAP gate should ensure that the axis-dependent local phase equal to $2n\pi$ (n = 1,2,3 …). The rest of the local phase shift is a duration-dependent correction from the Stark shift. This duration–dependent local phase can be calibrated by an echo sequence after initializing the two qubits in the same state, and can be eliminated by a virtual Z-gate.

Due to the difficulties in driving $Q_L$ in this device (see Supplementary Information Note 5), we have designed a quantum circuit to demonstrate the existence of the axis-dependent local phase using only single-qubit operations on $Q_R$, as illustrated in Fig. 3a. By inserting a wait time $\tau_{wait}$ between two consecutive SWAP gates, the axis-dependent local phase accumulated before the first SWAP gate is counteracted, and only the part that accumulated during the wait time remains. The two SWAP gates are inserted into a Hahn echo sequence to measure the single-qubit local phase accumulated on $Q_R$. We fix the idle time between the first single-qubit operation and the first SWAP gate and then observe the local phase acquired by $Q_R$ during $\tau_{wait}$. The spin-up-probability of $Q_R$ is read out, and the fitted oscillation frequency $f_{mp} = 7.90$ MHz is very close to the Zeeman energy difference at zero detuning $\Delta E_z(0) = 7.73$ MHz, which confirms the existence of the axis-dependent phase.

To calibrate the duration-dependent local phase, we use a $SWAP^2$ operation rather than the SWAP operation. For the $SWAP^2$ operation, which commutes with the single-qubit Z gates, the axis-dependent phase always vanishes, while the duration-dependent phase would be twice that of the SWAP operation. In Fig. 3b, we insert a $SWAP^2$ operation into a Hahn echo sequence and sweep the rotation-axis of the second $\frac{\pi}{2}$ rotation to measure the extra phase accompanied by the $SWAP^2$ operation. Compared to the measurement result without the $SWAP^2$ operation, the extra local phase is $0.65\,\pi$ and it should equal to $2[(\omega - E_z) + d\omega] * T$ (details in Supplementary Information Note 3), where $E_z$ and $\omega$ are the average Zeeman energies at the SWAP operation point and the single-qubit operation point (the symmetric point) respectively, and $d\omega$ is the Zeeman energy difference at the single-qubit operation point. From the energy spectroscopy, we extract the parameter $\omega = E_z = 19.8$ GHz, $d\omega = 7.73$ MHz and $T = 40.0$ ns. The local phase can be theoretically calculated as $0.620\,\pi$, which matches well with the experimental result. The difference can be attributed to the residual exchange coupling at the symmetric operation point and the miscalibration of the waveform.

**Truth Table and Gate Fidelity.** Benefiting from the high tunability of $J/|\Delta E_z|$, the spin-up probability of the two qubits $P_{|Q_L\rangle=|\uparrow\rangle}$ and $P_{|Q_R\rangle=|\uparrow\rangle}$ can be read out in the same sequence, allowing the observation of the non-classical correlation and the estimate of the SWAP gate fidelity. Recall that we measure the spin state of $Q_L$ directly via spin-selective tunneling, while spin of $Q_R$ is measured after shuttling the electron to $Q_L$. The fidelity of this spin shuttle process is characterized as 93.5% (see details in Supplementary Information Note 2). The complete measurement protocol including initialization, control, and readout is illustrated in Fig. 1d. In our device, the readout fidelity is 74.3% for $Q_L$ and 68.0% for $Q_R$ (see Supplementary Information Note 2), which is mainly limited by the thermal-broadening of the reservoir and the cross-talk in the single-shot readout [34].

To assess the fidelity of the SWAP gate, we first prepare the system in four different states $|\downarrow\downarrow\rangle, |\uparrow\downarrow\rangle, |\downarrow\uparrow\rangle$ and $|\uparrow\uparrow\rangle$ and then apply the exchange gate with pulse duration $\tau_S$. As $\tau_S$ increases, we observe the oscillation between $P_{|\uparrow\downarrow\rangle}$ and $P_{|\downarrow\uparrow\rangle}$ for the initial states $|\uparrow\downarrow\rangle$ and $|\downarrow\uparrow\rangle$, while $P_{|\downarrow\downarrow\rangle}$ and $P_{|\uparrow\uparrow\rangle}$ remain constant (Fig. 4a). With the correlation between the input and output states, the truth table of the SWAP gate and the $\sqrt{\text{SWAP}}$ gate can be constructed. The ideal and experimental truth tables of the SWAP gate and the $\sqrt{\text{SWAP}}$ gate are shown in Fig. 4b-c, with the initialization and readout errors removed from the experimental results. The lower bounds of the two-qubit gate fidelities are given as $F_{SWAP}^{min} = 82.2\% \pm 1.5\%$ and $F_{\sqrt{SWAP}}^{min} = 96.6\% \pm 2.5\%$, respectively (see Supplementary Information Note2). Here the gate fidelity is defined by the logical basis fidelity [32,58] extracted from the truth table, which is defined as $F_{gate} = \sum(E_{ij})/4$ and only includes the errors in the Z basis. $E$ is the truth table and $i, j$ are the indexes of the non-zero elements in $E$.

To analyze the effect of various error sources and estimate the actual gate fidelity, we simulate a two-qubit system with the parameters extracted from the energy spectrum and assume that it is dominated by quasi-static charge noise [59,60] on $\varepsilon$. The standard deviation $\sigma_\varepsilon$ of the charge noise can be acquired by reproducing the damping in the exchange oscillation decay. As given in Fig. S5, $\sigma_\varepsilon$ at different detuning $\varepsilon$ is approximately 4 GHz, and its stability confirms the correctness of our noise model. We then further verify our noise model by comparing the predicted visibility with the experimental data. We measure the exchange oscillation and truth table at various detuning $\varepsilon$. The visibility $A$ is extracted by fitting the envelope curve of the exchange oscillation with $P_{|\uparrow\downarrow\rangle} = A\exp(t/T_{2,SWAP})^\alpha + B$. The visibilities extracted from the experimental data and the simulation are given in Fig. 5a, and their trends match well in the detuning range relevant to the experiment setting. The deviation and large error bars at high detuning originate from the difficulty in fitting $\alpha$ and A well simultaneously when the states are close to the anti-crossing with the S (0, 2) state.

With the quasi-static noise model, we calculate the upper-limit to the gate fidelity of our SWAP gate. The simulated gate fidelity is calculated as $F_{gate} = tr(U^\dagger U_{ideal})$, where $U$ is the operator with errors and $U_{ideal}$ is the ideal gate operator. To estimate the infidelity caused by individual error sources, we add the finite ratio $J/|\Delta E_z|$, the residual exchange coupling $J_{off}$ and the charge noise one by one (see Supplementary Information Note 3). As shown in Fig. 5b, the infidelity considering only finite $J/|\Delta E_z|$ is strongly dependent on the detuning $\varepsilon$. This is evident because $J/|\Delta E_z|$ is a sensitive function of $\varepsilon$ and determines the rotation axis. The gate fidelity should approach 100% when $J/|\Delta E_z| \to \infty$ (the peak position in Fig. 5b), with the residual infidelity caused by the precision of the numerical calculation (see Supplementary Information Note 4). Another error source is the non-vanishing exchange coupling at the idle point. Normally, the exchange coupling at the symmetric operation point is difficult to turn off completely. The residual coupling $J_{off}$ is approximately 800 kHz in our experiments, which is included the simulation. As derived in Supplementary Information Note 4, $J_{off}$ changes the eigenstates at the symmetric operation point and the operation fidelity in the new basis is lower. The result in Fig. 5b shows that $J_{off}$ decreased the fidelity by 0.25%. Finally, we take the influence of charge noise into account. The result indicates that the gate infidelity impacted by the charge noise is only about 0.05% (see Supplementary Information Note 4). Compared to the infidelity introduced by the coherent error sources, the two-qubit SWAP gate fidelity is less influenced by the decoherence. Realizing a highly adjustable $\frac{J_{tr}}{dE_{tr}} = \frac{Jd\omega - |dE_z|J_{off}}{JJ_{off} + |dE_z|d\omega}$ is the core of demonstrating a high fidelity SWAP gate, where $J_{tr}$ and $dE_{tr}$ are the exchange coupling and the Zeeman energy difference in the new basis (see Supplementary Information Note 4). To realize the single-qubit gate and SWAP gate with a fidelity higher than 99% simultaneously, the ratio $\frac{J_{tr}}{dE_{tr}}$ must be tuned from lower than 0.1 to larger than 10 diabaticly. As shown in Fig. 5c, the tendency of the simulated gate fidelity fits well with the experimental result. We speculate that the actual gate fidelity should be higher than the estimated number from the experiment, due to the cross-talk error in the single-shot readout and the initialization error with chirped pulses. Hence, the experimental gate fidelities are only the lower bounds of the actual gate fidelities, and the latter ones should be between the simulation and the experiment results. It is also noteworthy that the experimental operation fidelity of the SWAP gate and the $\sqrt{SWAP}$ gate are quite different. We believe the discrepancy is due to readout and initialization errors, which causes nonlinear shift to the two operation fidelities respectively (see Supplementary Information Note 4).

**Conclusions.**

The next step in this study is to demonstrate a high-fidelity SWAP gate with the help of high-fidelity initialization and measurement. Beyond that demonstration, the natural progression of studies includes performing multiple SWAP operations in a larger structure of three or more qubits, and further improving the SWAP gate fidelity. When performing multiple SWAP gates, it may be beneficial to recall that while the native SWAP gate requires uniform external field, it does not need to be accompanied by single-qubit gates. The accumulated single-qubit phase errors can be corrected at the end of a SWAP sequence. As such micromagnets can be placed strategically and with simple designs, allowing better scalability. As for improving SWAP gate fidelity, recall that coherent errors such as finite $J/|\Delta E_z|$, $J_{off}$, and pulse errors are the dominant error sources for a SWAP gate. Pulse shape design could be an efficient approach to mitigate these coherent errors [36,61]. High circuit bandwidth and careful distortion calibration are also necessary to improve the gate fidelity. Last but not least, modifying the exchange coupling $J$ through barrier gate voltage (instead of detuning) could lead to a wider range of $J$ values, making the $J/\Delta E_z$ ratio tunable in a larger range [62].

Compared to the two-qubit CPhase gate and the CROT gate, the SWAP gate is realized through control of exchange interaction only, whose fidelity is in turn dominated by the charge noise on $J$. When a series of SWAP is performed, single-qubit phase correction is only needed at the end, making the whole operation more streamlined. When using the SWAP gate to manipulate or transfer the qubit states, the operation speed is typically faster than the two-qubit CPhase gate and the CROT gate due to the larger exchange coupling. In addition, in a quantum dot array, SWAP gates do not cause significant cross-talk and heating due to its very local nature, making it a more controllable tool in on-chip communication. Lastly, the demonstration of a potentially high-fidelity SWAP gate here shows a clear promise to engineer the Heisenberg Hamiltonian [63-65] in silicon for the purpose of quantum simulations.

In summary, we demonstrate the native SWAP gate for spin qubits in a gate-defined double quantum dot in isotopically purified $^{28}$Si. We implement a strategy to calibrate and eliminate local phase shifts induced during the SWAP gate. The truth tables are measured to give the operation fidelity, and the influences of various error sources are analyzed to estimate and further improve gate fidelity. Although the mis-calibrated SPAM error limits the extracted gate fidelity in our experiment, the demonstration indicates that a high-fidelity SWAP gate can be achieved even in the presence of a micromagnet, which can in turn facilitate on-chip quantum communication, and act as an essential two-qubit gate for quantum information processing. Future experiments will focus on reducing SPAM errors and optimizing the control pulse scheme, including engineering the pulse shape and tuning the exchange coupling with the barrier gate, so that we can realize a high-fidelity SWAP gate even with a larger $\Delta E_z$.

## Methods

**Device structure and fabrication.**

The DQD is formed underneath the two plunger gates PL and PR, and the reservoir is formed underneath the reservoir gate RG. The barrier gates BL and BM control the confined barrier between the reservoir, $Q_L$, and $Q_R$. The position of the SET is under gate S3. Three layers of 35, 55, and 65 nm aluminum electrodes overlap in a stack, with oxidized aluminum separating the different layers. A Ti/Co/Ti rectangular magnet is deposited on top of the device to provide the slanting Zeeman field for qubit manipulation with EDSR.

## Data availability

All the data that support the findings of this study are available from the corresponding author upon reasonable request.

**Acknowledgements**

This work was supported by the National Natural Science Foundation of China (Grants No. 12074368, 92165207, 12034018 and 92265113), the Innovation Program for Quantum Science and Technology (Grant No. 2021ZD0302300), the Anhui Province Natural Science Foundation (Grants No.

2108085J03), and the USTC Tang Scholarship. This work was partially carried out at the USTC Center for Micro and Nanoscale Research and Fabrication. H.-W. J. and X. H. acknowledge financial support by U.S. ARO through Grant No. W911NF2310016 and No. W911NF2310018, respectively.

## Author contributions

M. N. performed the bulk of measurement and data analysis with the help of S.-K. Z., X. X. and H.-O. L., and R.-L. M. fabricated the device. Z.-Z. K. and G.-L. W. supplied the purified silicon substrate. M. N. wrote the manuscript with inputs from other authors. X. H. provided theoretical support and G.-C.G. and H.-W. J. advised on experiments. C. W., A.-R. L., N. C. and W.-Z. L. contributed to the simulation and X. X. and G. C. polished the manuscript. H.-O. L. and G.-P. G. supervised the project. All the authors contributed to discussions.

## Competing interests

All authors declare that they have no competing interests.

# Figure captions

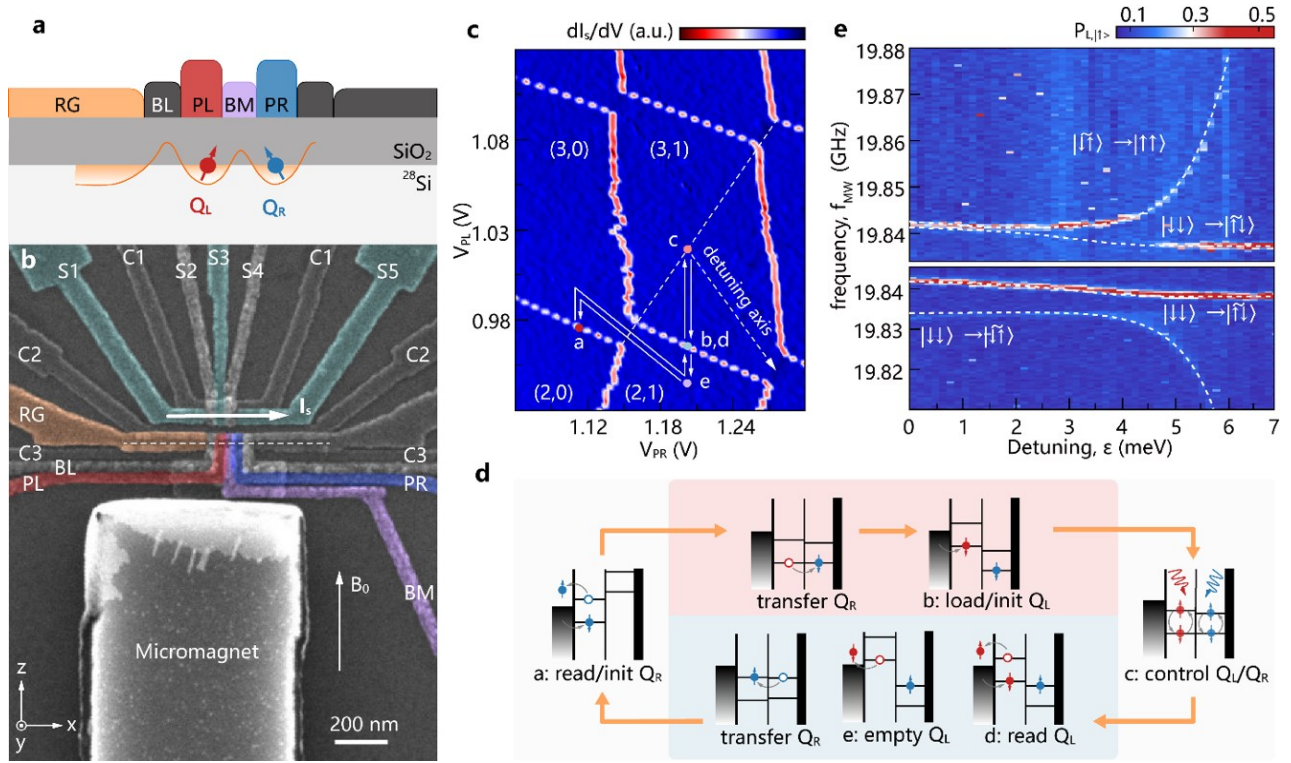

**Fig. 1. Two-qubit device, single-shot readout, and energy spectroscopy. a-b,** Schematic and false-color scanning microscope image of a device identical to the one measured in the experiment. The gate electrodes and the micro-magnet used in the experiment are indicated. The two qubits, $Q_L$ and $Q_R$, are located underneath the gates PL (red) and PR (blue), and the confinement barrier between them is formed underneath the gate electrode BM (purple). The single electron transistor (SET) is located under S3 and is turned by the SET gates S1-S5. The SET current $I_s$ is indicated by the white arrow. The reservoir underneath the gate RG (orange) supplies electrons to the quantum dots. The direction of the external magnetic field $B_0$ is shown by the white arrow in the right-down corner. Schematic **a** illustrates the profile map along the white dotted line. **c,** Charge stability diagram of the DQD formed underneath PL and PR. The number of electrons in the dots is represented by $(N_L, N_R)$. The sequential readout is achieved by pulsing from d to a. Points a to e indicate the initialization, control, and readout positions as depicted in **d**. **e,** Energy spectroscopy shows the spin-up probability of $Q_L$ versus the microwave frequency and the detuning $\varepsilon$ after initializing the system in $|\widetilde{\downarrow\uparrow}\rangle$ (top) and $|\downarrow\downarrow\rangle$ (bottom) states. We estimate $t_c = 1.5$ GHz and the Stark shift near the symmetric operation point to be 128.6

MHz/V. The lever arm along the detuning axis is 0.14 meV/mV, which is extracted from the magneto-spectroscopy. The white dashed curves are the numerical fitted results of the frequency resonance peaks and correspond to different transformation processes.

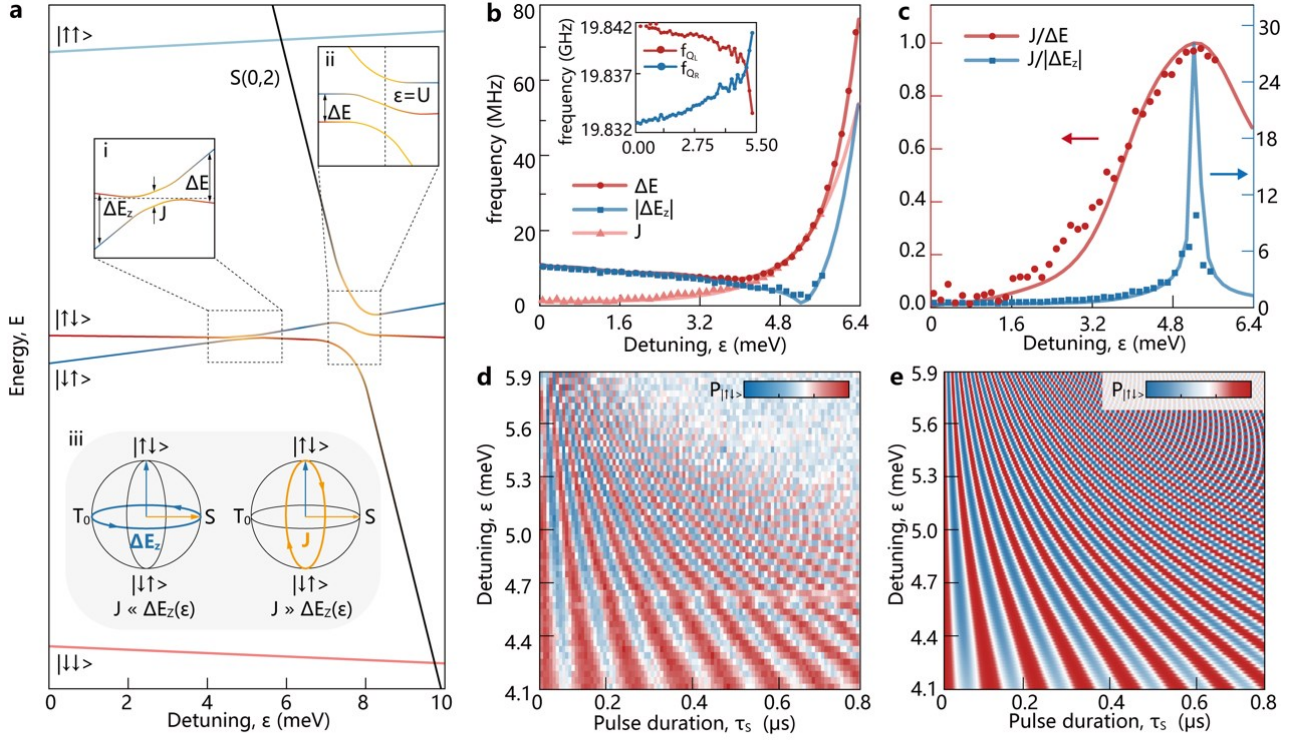

**Fig. 2. Energy-level structure. a,** Schematic of the energy-level structure of the two-electron spin states as functions of the detuning $\varepsilon$ between the (3,1) and (2,2) charge states. By exploiting the Stark shift, the energy levels of the two antiparallel spin states cross with each other (inset i). At the anticrossing between (3,1) and (2,2), the energy levels of the two antiparallel spin states shift by half of the exchange energy $J$ (inset ii). Around the detuning $\varepsilon = 0$, $J$ is small and the difference in Zeeman energy $\Delta E_z$ drives the rotation about the z-axis (blue arrow in inset iii). Near the two anti-crossings, the exchange energy $J$ is much larger than the difference in Zeeman energy between the two qubits, and the rotation about the x-axis (yellow arrow in inset iii) dominates the electron spin evolution interaction. **b,** $\Delta E_z$, $J$ and $\Delta E$ between the two mixing states $|\widetilde{\uparrow\downarrow}\rangle$ and $|\widetilde{\downarrow\uparrow}\rangle$ as functions of the detuning ε. Here, $\Delta E$ and $J$ are extracted from Fig. 1e. The exchange interaction is fitted with $J \propto e^{c\varepsilon}$, and $\Delta E_z$ is fitted with a fifth-order polynomial curve. $\Delta E_z = \sqrt{\Delta E^2 - J^2}$ is determined by $J$ and $\Delta E$. With the influence of the Stark shift, the single-qubit resonant frequencies as a function of $\varepsilon$ are given in the inset. The red and blue data correspond to resonant frequencies of $Q_L$ and $Q_R$, respectively. **c,** Ratio $J/\Delta E$ and the ratio $J/|\Delta E_z|$ as functions of the detuning $\varepsilon$. The red and blue solid lines are numerical results based on the fitted curve in **b**. **d,** The probability of $|\uparrow\downarrow\rangle$ as a function of $\varepsilon$ and the pulse duration $\tau_s$ after preparing the two-qubit system to $|\uparrow\downarrow\rangle$. The exchange oscillation fringe in **e** is the simulation of **d** obtained by solving the time-independent Schrödinger equation without dissipation.

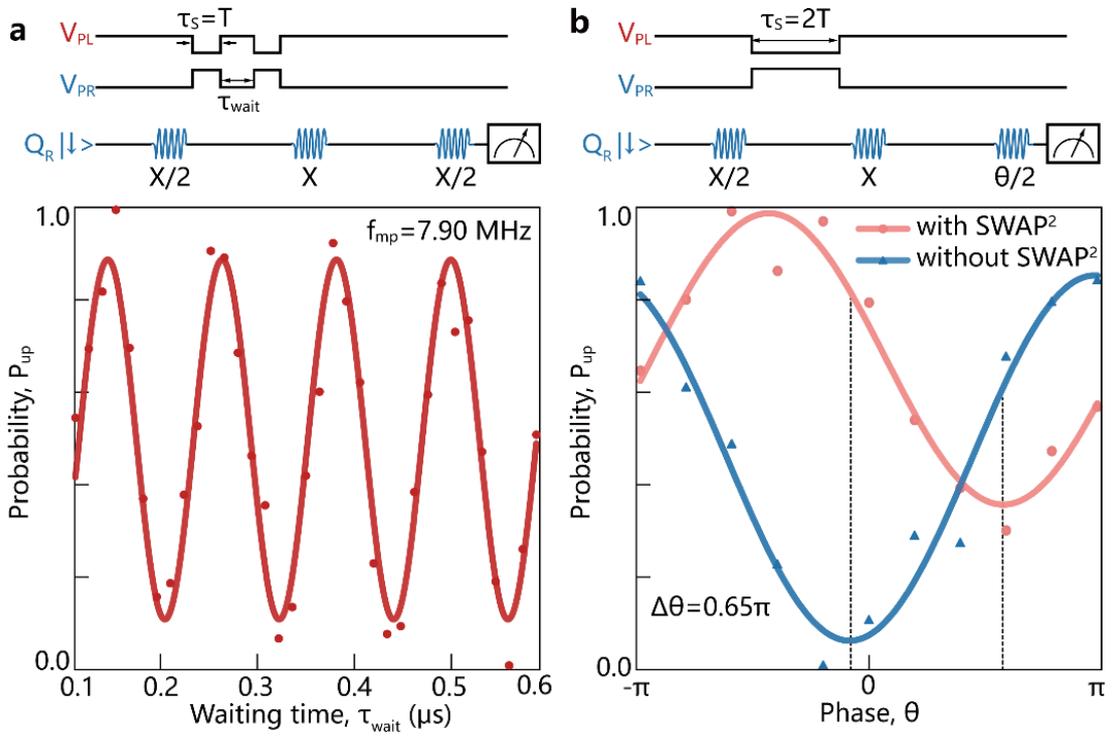

**Fig. 3. Local phase calibration. a,** The axis-dependent local phase accumulates as a function of waiting time $\tau_{wait}$. After initializing the qubits to the $|\downarrow\downarrow\rangle$ state, the diabatic pulse is inserted in an Hahn echo sequence, and the spin-up probability of $Q_R$ is read-out. Here, the pulse duration $\tau_s$ equals to the duration of a SWAP gate $T$. **b,** The duration-dependent local phase of a $SWAP^2$ operation is calibrated by sweeping the phase of the latter $\frac{\pi}{2}$ rotation in an echo sequence. Here the pulse duration $\tau_s$ equals to $2T$. The pink and blue data points correspond to the measurement results with and without the $SWAP^2$ operation respectively.

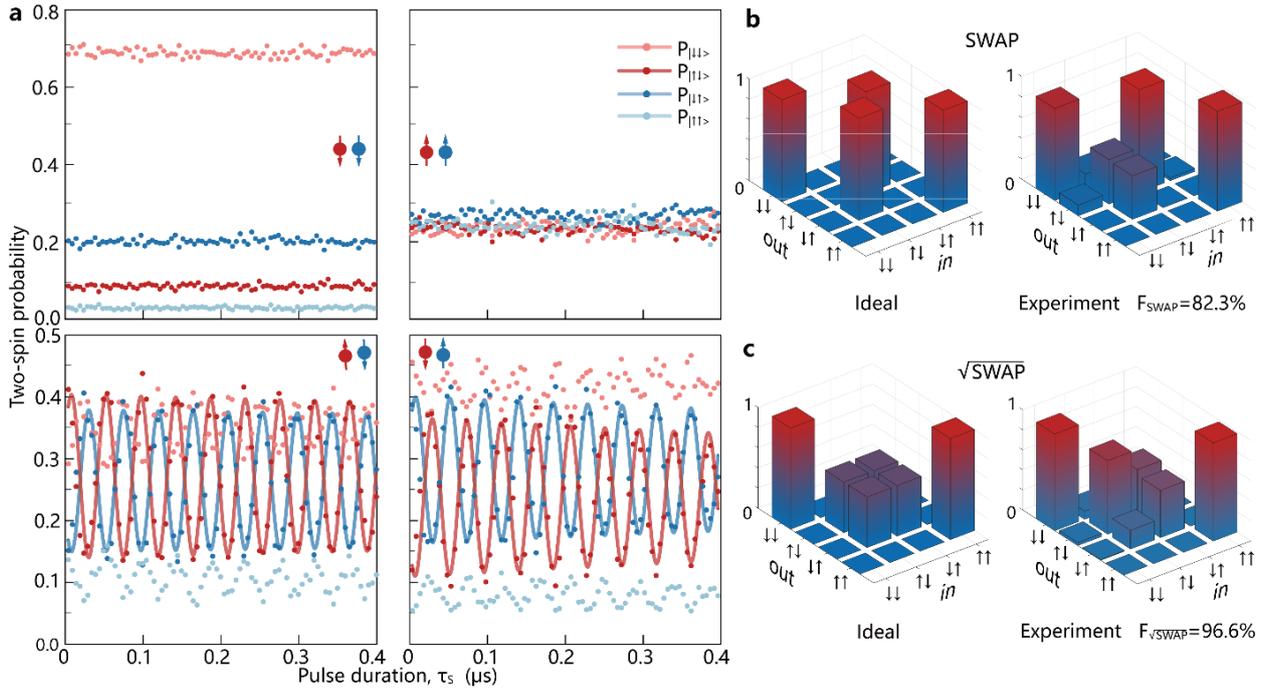

**Fig. 4. Coherent SWAP gate with the truth table. a,** The two-qubit probabilities as a function of $\tau_S$ with the initial state in $|\downarrow\downarrow\rangle$ $|\uparrow\downarrow\rangle$ $|\downarrow\uparrow\rangle$ and $|\uparrow\uparrow\rangle$, respectively. The initial states are illustrated in the corresponding panel. The solid lines fit to the oscillations of $P_{|\uparrow\downarrow\rangle}$ and $P_{|\downarrow\uparrow\rangle}$. **b-c,** Ideal and experimental truth tables of the SWAP and the $\sqrt{\text{SWAP}}$ gate. The experimental truth tables are extracted from **a**, where the SWAP gate duration is completed with a half period of exchange oscillation at $\tau_S \approx 25$ ns and the $\sqrt{\text{SWAP}}$ gate duration is $\tau_S \approx 12$ ns. The truth tables yield the lower bounds of the two-qubit gate fidelities as $F_{SWAP}^{min} = 82.2\% \pm 1.5\%$ and $F_{\sqrt{SWAP}}^{min} = 96.6\% \pm 2.5\%$, respectively.

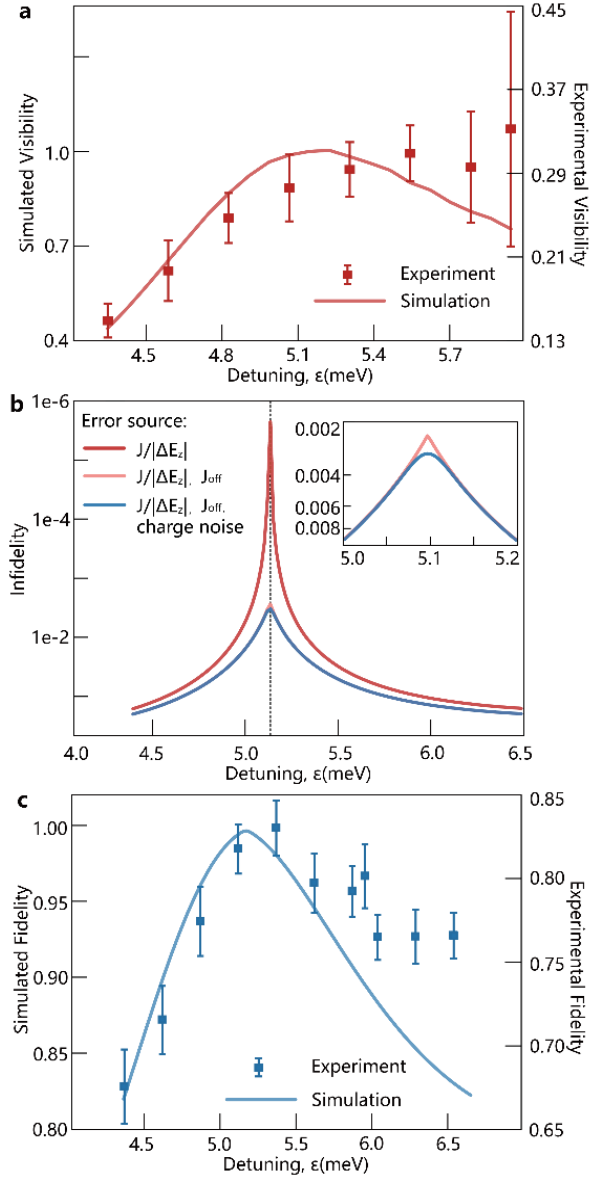

**Fig. 5. Noise and fidelity analysis. a,** The visibility of the exchange oscillation as a function of $\varepsilon$. The red dots are the experimental data, and the curve is simulated with $\sigma_\varepsilon = 4$ GHz and the parameters extracted from the energy spectroscopy. The visibility $A$ is extracted by fitting the envelope curve of the exchange oscillation with $P_{|\uparrow\downarrow\rangle} = A\exp(t/T_{2,SWAP})^\alpha + B$. **b,** The infidelity of the SWAP gate as a function of $\varepsilon$. The red curve indicates the simulated infidelity only accounting for the influence of $J/|\Delta E_z|$. The pink curve indicates the infidelity influenced by both $J/|\Delta E_z|$ and $J_{off}$. The blue curve indicates the infidelity influenced by $J/|\Delta E_z|$, $J_{off}$, and charge noise. The dashed line indicates the maximum value of the infidelities. **c,** The gate fidelity as a function of $\varepsilon$. The data points calculated from the truth table are influenced by the initialization and readout error hence giving the lower bounds of the gate fidelities.